\documentclass[12pt,preprint]{aastex}

\shorttitle{Tidal tails around globular clusters}

\shortauthors{}

\begin{document}


\title{Tidal tails around globular clusters.\\
Are they a good tracer of  cluster orbits?}

\author{M. Montuori}
\email{montuorm@roma1.infn.it}
\affil{CNR, Roma, Italy}
\author{R. Capuzzo-Dolcetta}
\email{roberto.capuzzodolcetta@uniroma1.it}
\affil{Dep. of Physics, University of Roma 'La Sapienza', Roma, Italy}
\author{P. Di Matteo}
\email{paola.dimatteo@obspm.fr}
\affil{LERMA, Observatoire de Paris, Paris, France}
\author{A. Lepinette}
\email{lepinette@inta.es}
\affil{Centro de Astrobiologia, CSIC/INTA, Torrejon de Ardoz, Spain}
\author{P. Miocchi}
\email{miocchi@uniroma1.it}
\affil{INAF, Osservatorio Astronomico di Collurania, Teramo, Italy}


\begin{abstract}
We present the results of detailed N-body  simulations
of clusters moving in a realistic Milky Way (MW) potential.
The strong interaction with the bulge and the disk of the
Galaxy leads to the formation of tidal tails, emanating from opposite
sides of the cluster. 
Some characteristic features in the morphology and orientation of these
streams are recognized and intepreted.
The tails have a complex morphology, in particular
when the cluster approaches its apogalacticon, showing multiple ``arms'' in
remarkable similarity to the structures observed around NGC 288 and Willman 1. \\
 Actually, the tails are generally good tracers of the
cluster path quite far from the cluster center ($>7$--$8$ tidal radii),
while on the smaller scale they
are mainly pointing in the direction of the Galaxy center.
In particular, the orientation of the inner part of the tails is highly
correlated to the cluster orbital phase and to the local orbital angular acceleration.
This implies that, in general, the orbital path cannot be estimated directly from the orientation
of the tails, unless a sufficient large field around the cluster is available.
\end{abstract}




\section{Introduction}
It is commonly accepted that the present globular cluster (GC) population
in the Milky Way represents the survivor of an initially more numerous one, depauperated by many
disruptive processes \citep{mw97a,mw97b,fz01}.\\
The observational results obtained in the last decade have clearly confirmed the
role played by the Milky Way environment on the GC evolution. Up to now, about
$30$ galactic GCs show evidences of star depletion due to the tides caused by
the field of the Galaxy. \\ The first evidence of the existence of
tails surrounding GCs were achieved by \citet{grill95}. Using colour-magnitude
selected star counts in a dozen of galactic GCs, these authors
showed that in the outer parts the stellar surface
density profiles exceeded significantly the prediction of King models,
extending also outside the tidal radius. They identified this
surrounding material as  made up of stars in the act of being tidally stripped
from the cluster field and pointed out the importance of defining the
large-scale distribution of extra-tidal stars on the sky to obtain constraints
and traces for the GC orbits. More recently, other works confirmed and enlarged
Grillmair et al.'s findings \citep{ls97, testa00, lmc00, sieg00, lee03} giving evidence
of the existence of many GCs surrounded by haloes or tails.
For M92, \citet{lee03} found, also, that the extratidal material is not randomly distributed,
being the density profile in the tails shallower for bright stars than for fainter ones.\\
This was the state of the art until the spectacular findings of two tidal tails
emanating from the outer part of the Palomar 5 globular cluster and covering an
arc of 10 degrees on the sky, corresponding to a (projected) length of 4 kpc at
the distance of the cluster \citep{oden01, oden02, oden03}, obtained in the
framework of the Sloan Digital Sky Survey (see also \emph{http://www.sdss.org}).
The stellar mass in the tails of this sparse, low-mass, halo cluster (with an
estimated concentration parameter $c=0.7$) adds up to 1.2
times the mass of stars in the cluster, estimated in the range between
$4.5\times 10^3 M_{\odot}$ and $6\times 10^3 M_{\odot}$.
More recently, \citet{grill06}, still using SDSS data, 
have detected a continuation of Pal 5's trailing tidal 
stream out to almost 19 degrees from the cluster. 
Combining this with the already known southern tail of 
Pal 5 yields a stream some $9 $ kpc long on the sky.\\
Substantial tidal streams have recently been found 
associated with another low-mass and low-concentration 
GC in the SDSS area: NGC 5466. 
For this cluster \citet{belok06} reported giant 
tails extended for about 1 kpc in 
length and \citet{grillj06}, still using SDSS data, 
suggested that a $\approx 13$ kpc tidal stream of stars,
extending from Bootes to Ursa Major, could also be associated with this system. \\
Together with the so-called Sagittarius stream
\citep{mateo98,yanny00,mart01,maj03,mart04}, which emerges from a dwarf galaxy
that is currently being accreted by the Milky Way, Palomar 5 and NGC 5466
represent outstanding examples of ongoing tidal erosion of  stellar systems in
the Milky Way, being, up to now, also the only two globulars known for which
such  extended stream-like structures have been detected in the Galactic halo.
\\

One of the first numerical investigations of the role played by a galactic tidal
field on spherical stellar systems was that of \citet{keenan75}, who studied the
effect of realistic, time-varying tidal fields on the stellar orbits in a star
cluster.  They numerically integrated the equations of motion of
three-bodies in models of spherically symmetric clusters which, in turn,
move in eccentric orbits in the field of a model galaxy.
One of the main conclusions of this work, which extended previous investigations
made by \citet{king62}, was that star clusters rotating in a retrograde sense are 
more stable in a tidal field than clusters with either direct rotation or no 
rotation due to the contribution of the Coriolis acceleration, acting in the 
same direction as the gravitational attraction for 
retrograde motion.\\
More recent works on weak tidal encounters based on a Fokker-Planck approach
\citep{ohlin92, leegood95} and self-consistent N-body techniques \citep{grill98} 
confirmed that the interaction with external tidal field, combined with 
two-body relaxation in the core of the cluster and following replenishment of 
stars near the tidal radius, causes a flow of stars away
from the cluster. The stripped stars remain in the vicinity of the cluster for
several orbital periods, migrating either ahead of the cluster or falling
behind, giving rise to a slow growth of the tidal tails. \\
A semi-analytic study of the development of tidal streams in galactic satellites
was done by \citet{john98}, who gave estimates for the rate of growth of 
tidal tails.
The effects of a realistic galactic tidal field (including both bulge, halo and
disk) on GCs were investigated few years later by \citet{clm99}.
The main findings of the  work  were the following: stars escaped from the
system go to populate two giant tidal tails along the cluster orbit; these tails 
present substructures, or clumps,  attributed to strong shocks suffered by the 
cluster, and are preferentially formed by low mass stars.\\
\citet{yimlee02} performed N-body simulations of GCs orbiting in a two-component
galaxy model (with bulge and halo and no disk), using the direct-summation 
NBODY6 code \citep{aarseth99} and focusing their attention, in particular, to 
the correlation between tidal tail elongation (described by mean of a 'position 
angle', defined as the angle between the direction of the tail and the galactic
center direction) and the cluster orbit. They found that, on circular orbits, 
tidal tails mantain an almost constant position angle ($\sim 60^{\circ}$),
while GCs on non circular orbits show a variation of the position
angle, according to orbital path and phase. The position angle increases when 
the cluster heads for perigalacticon. 
On the other hand, it tends to decreases when the cluster heads for apogalacticon. 


Finally, some authors investigated also the 
dynamical evolution of some globular clusters in the tidal field of the 
Galaxy. In this context \citet{dehnen04} modelled the disruption of
the globular cluster Pal 5 by galactic tides. Pal 5 is remarkable not only for its extended and massive tidal tails, but also for 
its very low mass and velocity dispersion. In order to understand these 
extreme properties, they performed many simulations aiming at reproducing the Pal 5 evolution along its orbit across the Milky Way. They explained the 
very large size of Pal 5 as the result of an expansion following the 
heating induced by the last strong disk shock about 150 Myr ago.
The clumpy substructures detected in the tidal tails of Pal 5 are not reproduced 
in their simulations, so that they argued that these overdensities have been probably caused by interaction with Galactic substructures, such as giant 
molecular clouds, spiral arms, and dark  matter clumps, which were not condsidered in their modeling.
These simulations also predict the destruction of Pal 5 at its next disk
crossing in  about 110 Myr, suggesting that many more similar systems have once populated the inner 
parts of the Milky Way but have been transformed into debris streams by 
the Galactic tidal field.  \\
In this context, it may be interesting mentioning  the recent numerical work
devoted to the study of smaller size systems (open clusters) in the MW tidal
field \citep{chum06}, which confirms already known results in the case of the 
external part of GC tidal tails) about the alignement of stars of the tidal 
stream around a common orbit in the external field \citep{grill98, clm99, cdm05}.

In the above sketched theoretical and observational background, this work --which is
inserted in  a wider study on the dynamics of globular clusters in external 
tidal fields  \citep{cdm05, dm04, mioc, mioc2}-- is devoted to clarifying the connection
among tidal tails and cluster orbit. We will
describe and discuss the mechanisms that determines the tails morphology and how
they depend on cluster trajectory and orbital phase. \\
For this purpose, we performed detailed $N-$body simulations (with $N=10^5$) of 
GCs moving in a realistic three-components (bulge, disk and halo) Milky Way 
potential. \\
The paper is organized as follows: in Sect.\ref{Cluster} 
the Galaxy (Sect.\ref{Galaxy}) and cluster (Sect.\ref{Cluster1}) 
models adopted are presented, so as the numerical approach used; 
in Sect.\ref{results} we deal with the main results of our work, showing the 
formation and development of tidal tails around the cluster (Sect. \ref{tails}), 
giving a qualitative approach for describing the tail morphology 
(Sect.\ref{qualit}), presenting the  numerical procedure adopted to fit tails 
direction (Sect.\ref{fit}) and, finally, discussing the tail-orbit alignement 
and its dependence on the orbital phase. 
In Sect.\ref{concl}, all the results are summarized and discussed. 

\section{Models and methods}\label{Cluster}
\subsection{Galaxy model}\label{Galaxy}
The model adopted for the Galactic mass distribution is that of
\citet{allsant91}. It consists of a three component system: a spherical central
bulge and a flattened disk, both of \citet{Miyamoto-Nagai} form, plus a massive
spherical halo.
The gravitational potential is time indipendent, axysimmetric and given in an analytical form which is continuous together with its spatial
derivatives.\\
Choosing a reference frame where the ({\it x,y}) plane coincides with the MW equatorial plane, the 
three components of the potential have, in cylindrical coordinates, the form:
\begin{eqnarray}
\Phi_B(\omega,z)&=&-\frac{GM_B}{\sqrt{R^2+z^2+{b_B}^2}}\\
\Phi_D(\omega,z)&=&-\frac{GM_D}{\sqrt{R^2+\left(a_D+\sqrt{z^2+{b_B}^2}
\right)^2}}\\
\Phi_H(r)&=&-\frac{GM(r)}{r}-
\frac{GM_H}{1.02a_H}\left[-\frac{1.02}{1+(\frac{r}{a_H})^{1.02}}+ln
\left(1+(\frac{r}{a_H})^{1.02}\right)\right]^{100}_r,\\
\end{eqnarray}
where square brackets indicate the difference of the function evaluated at the 100 
kpc and the generic $r=\sqrt{R^2+z^2}$ extremes. 
The parameters in the formulas above are listed in Table \ref{tbl1}.

\subsection{Cluster model and numerical method}\label{Cluster1}

The total initial mass of the cluster was chosen $M_{GC}=4.7 \times 10^5$ M$_{\odot}$,
i.e. a value compatible with masses of galactic globular clusters lying inside 4.5 kpc from the MW center (see \citet{har96} and \citet{pry93}).
The stellar mass spectrum of our cluster initial model
was chosen as a Kroupa IMF sampled in the
range $0.1M_{\odot}-20M_{\odot}$ and "evolved" (in the sense of
accounting for  mass loss on the base of stellar evolution,
according to \citet{scl97,dom99}) up to $3$ Gyr (which corresponds to our
assumed cluster age). In this interval of time all masses greater than $1.2$
M$_\odot$ go into the $0.5M_{\odot}-1.2M_{\odot}$ range.
As in \citet{cdm05}, we sampled this mass function into 12 mass classes, equally spaced in a
linear scale, whose space and velocity distribution were evaluated by
the adoption of a multimass King distribution \citep{king66,costa76}.\\
Obviously, the choice of $N=10^5$ as total number of cluster stars, together with
the given  value of $M_{GC}$ and of the above described mass function
implied a rescaling of the star masses.
Finally, to investigate the role played by the degree of cluster concentration,
we considered clusters with two different values for the King concentration
parameter $c$ (listed in Table \ref{tbl2}).    \\

\begin{figure}
\begin{center}
\includegraphics[angle=0,scale=0.7]{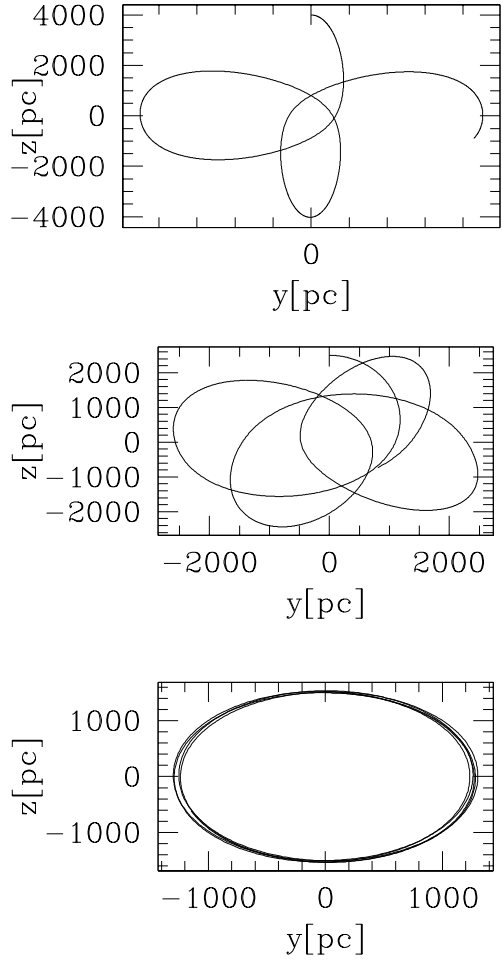}
\caption{
Orbits of a point mass with the initial position and velocity
given in Table \ref{tbl3} in the MW potential (see sect. \ref{Galaxy}). 
Top panel: orbit I; central panel: orbit II; 
bottom panel: orbit III.
\label{orbits}}
\end{center}
\end{figure}

The clusters move on the $y-z$ coordinate plane (the ($x,y$) plane corresponds to
the Galactic disk), along orbits of different eccentricity, defined as
\begin{equation}
e=\frac{r_a-r_p}{r_a+r_p}, \label{ecc}
\end{equation}
being $r_p$ and $r_a$, respectively, the GC pericentric and apocentric orbital
distances. See Table \ref{tbl3} for GC orbital parameters and Fig.\ref{orbits}
for a plot of the different simulated orbits.

All the simulations were performed by means of the TreeATD code,
developed by two of us \citep{bib2} which, as the original code by \cite{bh}, 
relies upon a tree-algorithm for the gravitational force evaluation on 
the large scale and on a direct summation on the small scale. 
The code was parallelized to run on high performance computers (via MPI routines)
employing an original parallelization approach (see again \citep[see]{bib2}).
The time-integration of the `particles' trajectories is performed by a
leap-frog algorithm that uses individual and variable time-steps
according to the block-time scheme \citep{aars,bib5}.
See \citet{mioc} for further details.
The only constant of motion that can be significantly used to check
the quality of the orbital time--integration is the total energy of the
cluster, i.e. including the contribution of the external field
to the GC potential energy. We saw that the upper bound of the 
relative error in the energy conservation,
$\Delta E/E$, is $10^{-4}$ over the whole simulation time. 
This is no more than one order of magnitude worse than the
error we got in a comparison simulation for the same GC
in abscence of external field.

\section{Results}\label{results}
\subsection{Formation and evolution of tidal tails}\label{tails}
\begin{figure}
\includegraphics[angle=0,scale=0.7]{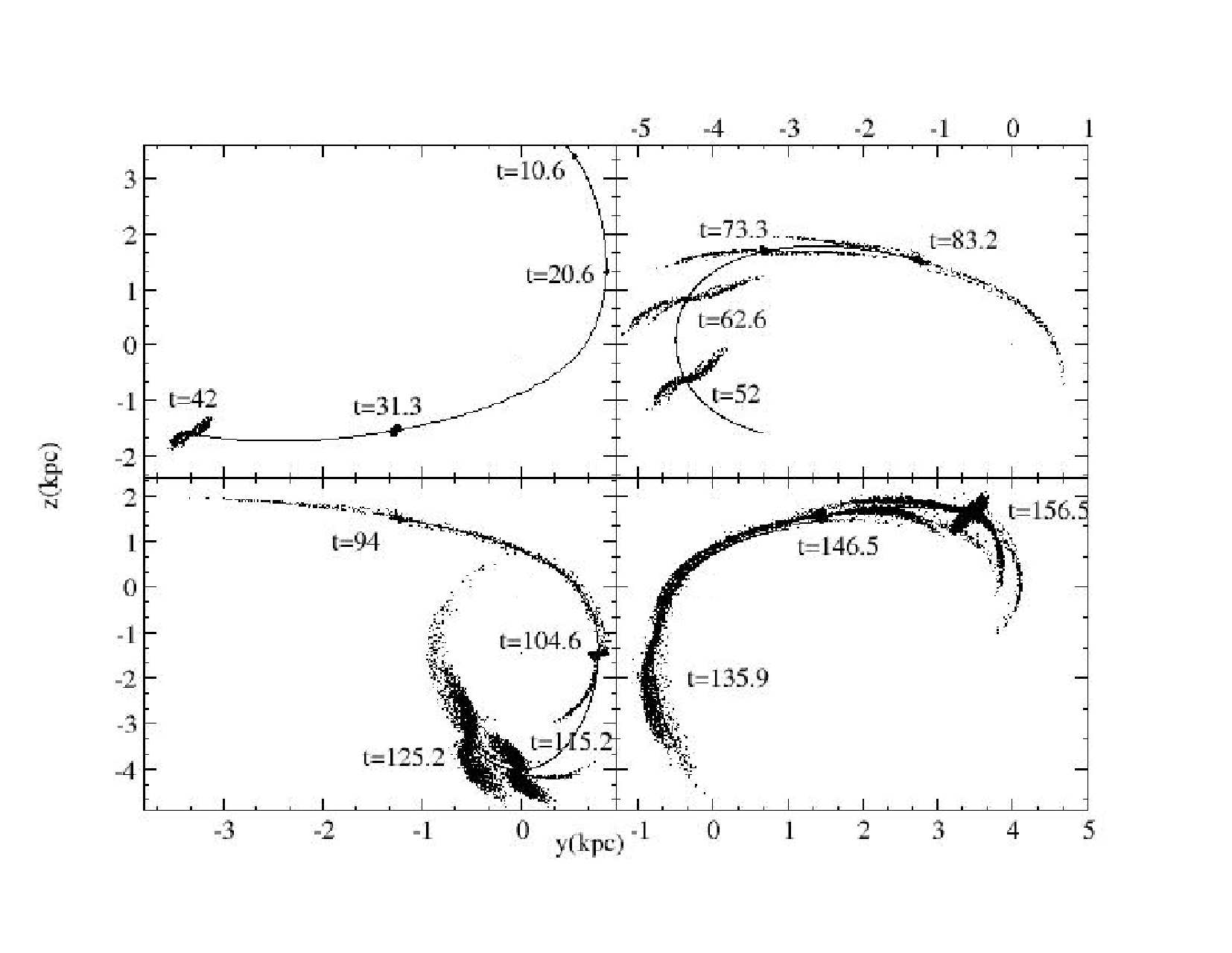}
\caption{
Some snapshots of the GC moving along the orbit I (see Table \ref{tbl3}
for the orbital parameters). 
One time unit corresponds to about $1$ Myr.
\label{orbitaI}}
\end{figure}

\begin{figure}
\includegraphics[angle=0,scale=0.7]{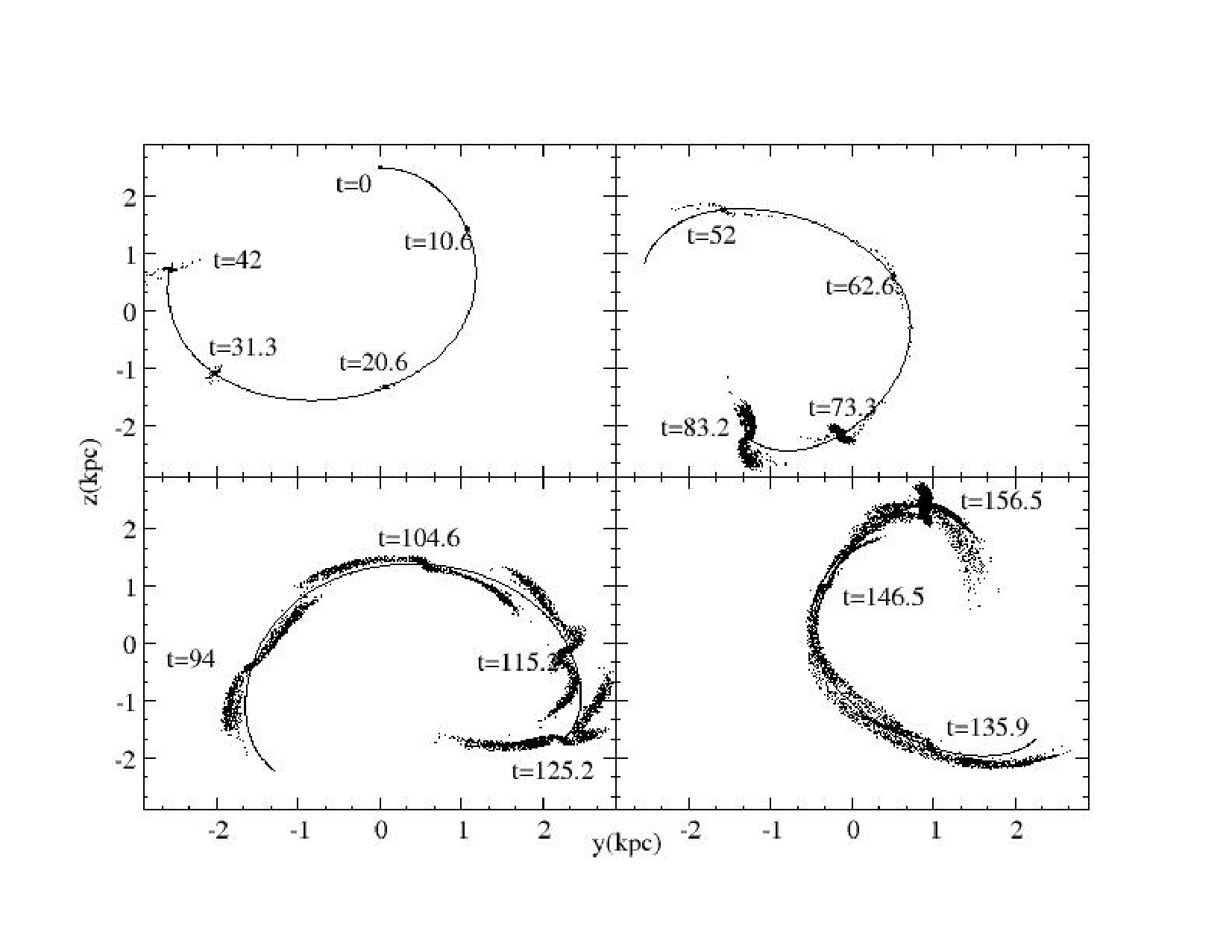}
\caption{Some snapshots of the GC moving along the orbit II (see Table \ref{tbl3}
for the orbital parameters). 
One time unit corresponds to about $1$ Myr.
\label{orbitaII}}
\end{figure}

\begin{figure}
\includegraphics[angle=0,scale=0.7]{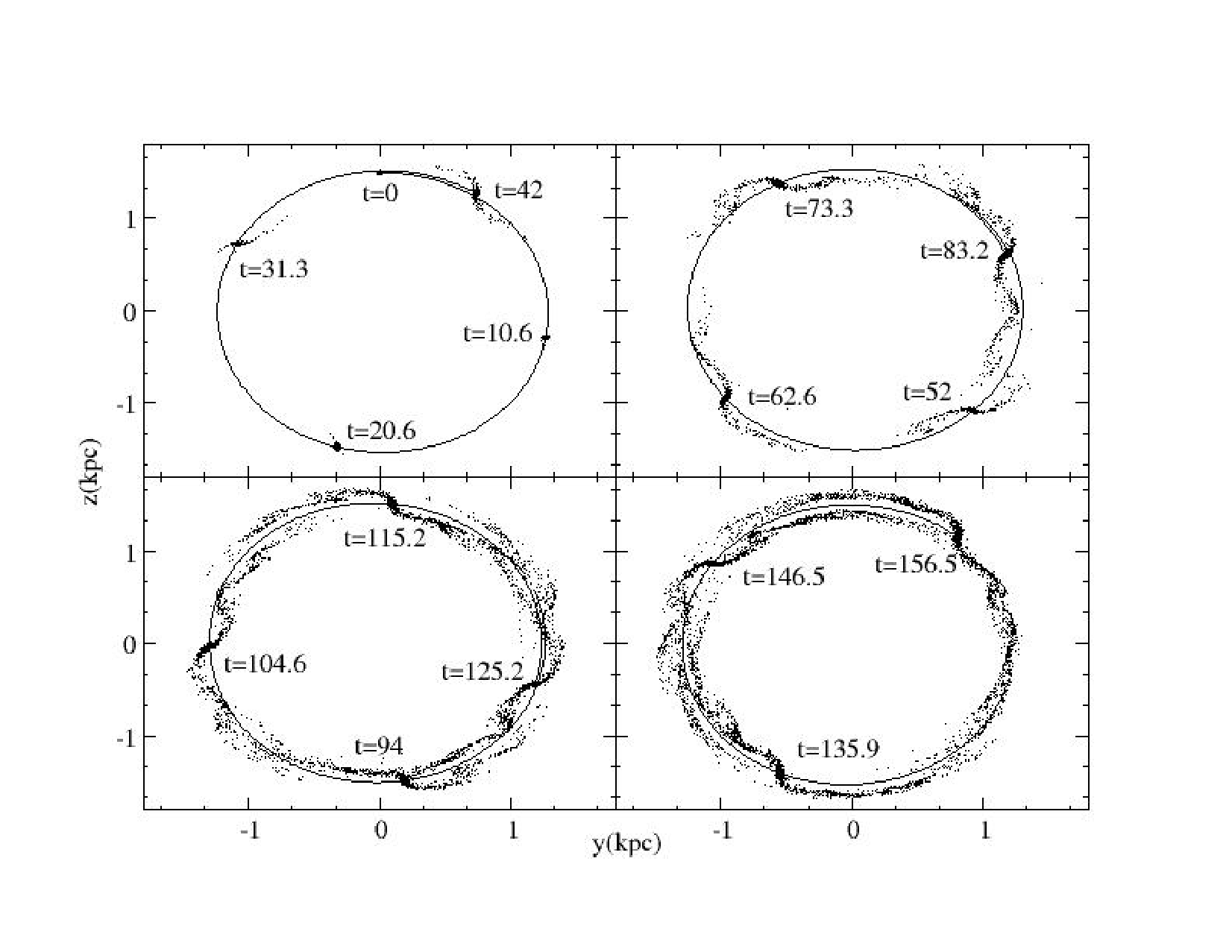}
\caption{Some snapshots of the GC moving along the orbit III (see Table
\ref{tbl3} for the orbital parameters). 
One time unit corresponds to about $1$ Myr.
\label{orbitaIII}}
\end{figure}

In Figs. \ref{orbitaI},\ref{orbitaII} and \ref{orbitaIII} the formation and subsequent
evolution of tidal tails around the GC is shown, for all the simulations
performed in the case of the least concentrated of the two GC models
(the one with $c=0.81$ in Table \ref{tbl2}). In the case of the more 
concentrated initial model cluster, the development of the tidal tails
shows the same time dependence being just slightly less populated.\\
After about $5\times10^7$ yr, tidal tails have clearly formed. They continuously
accrete by the stars leaving the cluster, so that after $1.6\times 10^8$ yr, in the
case of the quasi-circular orbit (orbit III), they are elongated for more than
1.5 kpc each. 
In general, the extention of the two cluster tails depends on the
velocity of the cluster and its variation along the orbit.
For the quasi-circular orbit (orbit III) this velocity is almost constant and the two tails
have nearly the same length. For more eccentric orbits (orbit I and II), the tail extending between
the GC and the pericenter is more elongated than the opposite tail, because the stars
in the latter have smaller velocities than those belonging to the former one.
In any case, the tail that precedes the cluster
always extends slightly below the orbit, whereas the trailing one lies slightly
above the orbit in agreement with what was observed for Pal 5.
This feature can be explained considering the role played by the Coriolis 
acceleration, as we will see later.

As regards the tails orientation, these figures (Figs
\ref{orbitaI},\ref{orbitaII},\ref{orbitaIII}) clearly show
the alignment of the outer part of the tails (at
distances from the GC center $>7$--$8$ tidal radii) with the orbit.
On the contrary, the alignment of the inner part is strongly correlated to both
the orbit eccentricity and the GC location. Notice, indeed, that the inner
tails are roughly aligned with the orbital path only when the cluster is 
near the perigalacticon of the more eccentric orbits (I and II, respectively
shown in Fig.\ref{orbitaI} and Fig.\ref{orbitaII}).
This confirms previous results in \citet{cdm05}, where analogous features
were found for GCs movings in a triaxial potential.

It is also worth noting the peculiar morphology in the streams when the
cluster is approaching the
apocenter in the eccentric orbits (orbits I and II): tidal
tails divide into ``multiple arms'', two nearly elongated along the GC path, while
the other two point to the galactic center (and anticenter) direction.
At this regard, the tails observed around NGC288 \citep{lmc00}
have a similar complex morphology, showing three different ``arms''.
Interestingly, this cluster has been estimated to be near its
apogalacticon \citep{dinescu} in agreement with our results.
Also the galactic satellite Willman 1 shows multi-directional stellar tails \citep{wil},
which according to our results, could represent evidence that 
this system is approaching the apocenter on an
eccentric orbit.

\subsection{Tidal tails elongation: a qualitative description}\label{qualit}
\begin{figure}
\includegraphics[angle=270,scale=0.65]{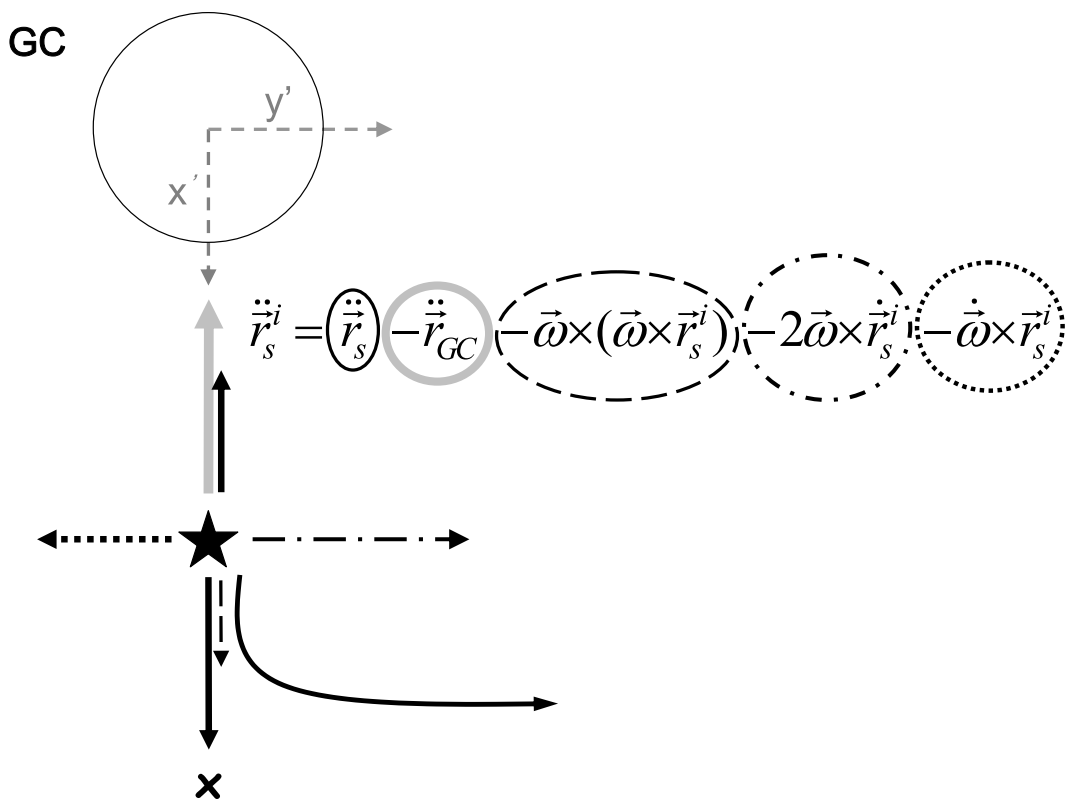}
\caption{Interpretation of the S-shape of the inner tidal tail around a
globular cluster. The different terms in the rhs of the displayed equation
(Eq. 7) are represented in the plot as arrows of different line styles.
Note that the last term in the equation is here plotted antiparallel
to the Coriolis term  as it occurs when
the GC moves from pericenter to apocenter (see text). The Galactic potential
(included in the first and second term) is assumed, for simplicity,
spherical. The cross in the lower part of the figure represents the Galaxy center.
\label{coriolis}}
\end{figure}

As anticipated in \citet{dm04}, the simplified scheme of the formation and shape of tidal tails
around GCs can be understood by the motion of a star 
escaping from a cluster moving in a spherical potential. Let us consider a rotating
reference frame with the origin in the cluster center-of-mass, the $(x',y')$
plane coinciding with the orbital plane and $x'$ pointing to the galactic center.
The equation of motion of the $i_{th}$ star belonging to the cluster
that moves around the galaxy center with variable angular
velocity $\mbox{\boldmath{$\omega$}}$ is (see also Appendix \ref{app} and Fig.\ref{coriolis}):
\begin{equation}
\ddot{\bf r}'_i=\ddot{\bf r}_i-\ddot{\bf r}_{GC}-2 \mbox{\boldmath{$\omega$}}
\times \dot{\bf r}'_i-\mbox{\boldmath{$\omega$}} \times
\left(\mbox{\boldmath{$\omega$}} \times \bf
{r}'_i\right)-\dot{\mbox{\boldmath{$\omega$}}}\times \bf {r}'_i, 
\label{equat}
\end{equation}
where ${\bf r}'_i$ is the position vector in the
rotating frame, while ${\bf r}_i$ and the position vector of the GC center-of-mass$, 
{\bf r}_{GC}$, refer to  the inertial frame. 
For a star escaping through
one of the unstable Lagrangian points along the $x'$ axis, 
the first, second and fourth terms on the
right-hand side of Eq. \ref{equat} are directed  
along the $x'$ direction, while the third term (the Coriolis acceleration)
and the fifth term are along  $y'$.
The latter is parallel to the Coriolis term when $|\mbox{\boldmath{$\omega$}}|$ increases
 (i.e. moving from the apogalacticon to perigalacticon) and antiparallel when
$|\mbox{\boldmath{$\omega$}}|$ decreases (moving from the perigalacticon to
the apogalacticon).
Consequently, these latter two terms ere responsible of the
initial deviation of the tails from the radial direction and of the formation 
of the S-shape profile \citep[see][]{cdm05}).

\subsection{The tail direction along the orbit}\label{fit}
As a reliable quantitative definition of the tail orientation, we used the direction
of the eigenvector corresponding to the greatest
eigenvalue\footnote{One of the three eigenvectors of the inertia
tensor is parallel to the $z$-axis, while the other two lye in the
($x,y$) plane. Among these latter two, the eigenvector associated to the greatest eigenvalue 
gives the direction of maximum elongation of the stellar stream.} of the inertia tensor of 
a suitable portion of the cluster. More precisely, we considered the 
stars contained in an annulus centered at the cluster
center-of-density (CD), 
as defined by \citet{ch85}, and
extending from $\approx 30$ pc up to a distance of $7$--$8$ tidal radii ($\sim 170$ pc).
This region includes, indeed, the tail portion we call "inner" part of the
tails; this part is that usually detected in the observations.
Note that in the study of \citet{yimlee02} the direction of the tails was determined by eye, 
without any mathematically defined procedure.\\

For each configuration of the system:
\begin{itemize}
\item we evaluated the GC center-of-density;
\item we selected all the stars distributed between $30$ pc and $170$ pc from the cluster CD
(excluding stars lying inside $r<30$ pc, because this region corresponds to the spherical
part of the system);
\item we evaluated the inertia tensor and the corresponding eigenvalues 
and eigenvectors of the selected stars;
\item we determined the eigenvector ${\bf u}$ corresponding to the maximum
eigenvalue.
\end{itemize}
 This eigenvector is parallel to the direction of the two tails having opposite
orientation. In the following, we always refer
to the tail internal to the cluster orbit.

The evolution of the tidal tail orientation is shown in Figs. \ref{tail-radI},
\ref{tail-radII}, \ref{tail-radIII}.\\
These figures show the GC orbit and some kinematic quantities plotted 
as functions of time.
From top to bottom we plot: the GC orbit (first panel), the distance $r$ of the cluster from the
Galaxy center (second panel), the magnitude of the angular velocity vector
$\mbox{\boldmath{$\omega$}}$ (third panel), the time derivative
of $|\mbox{\boldmath{$\omega$}}|$ (fourth panel).
Finally, the fifth panel shows the evolution with time of two quantities:
the angle

\begin{equation}
\alpha\equiv
\cos^{-1}\left({\frac{ {\bf u}  \cdot  {\bf r_{GC}}}{ |{\bf u}||{\bf r_{GC}}| }}\right)
\end{equation}
between the tail direction and that of the galactic center and the
angle 
\begin{equation}
\beta \equiv \cos^{-1}\left({\frac { {\bf u}\cdot \dot{\bf r}_{GC}} {|{\bf u}|
|\dot{\bf r}_{GC}|}}\right)
\end{equation}
between the tail direction and the GC velocity vector (the angles assume values
in the $[0^\circ,180^\circ[$ interval).\\

\subsubsection{Tail-Galaxy center alignement}\label{tailgal}
\begin{figure}
\includegraphics[angle=270,scale=0.7]{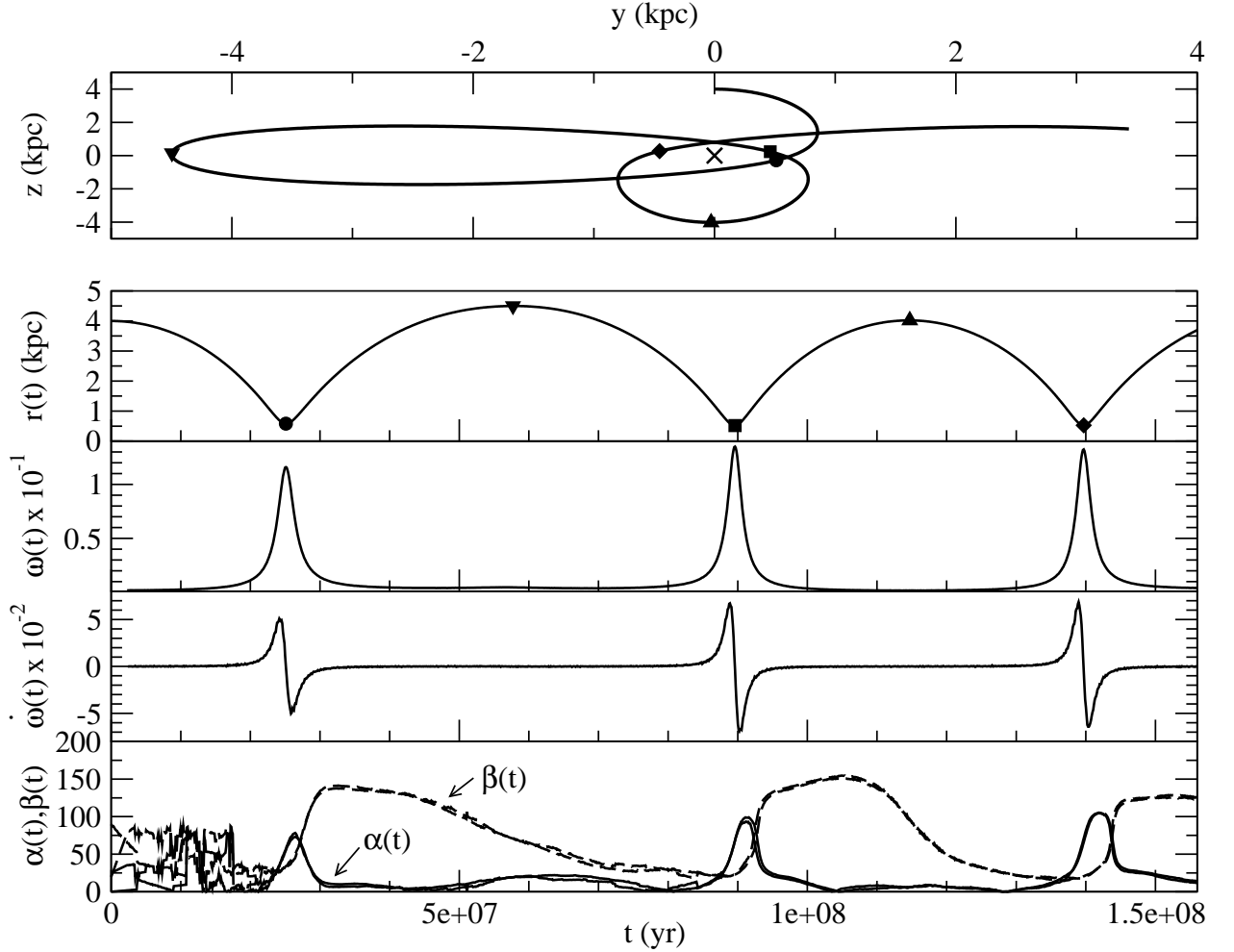}
\caption{
Tails direction for the GC moving on orbit I. From top to the bottom: 
Plot of the GC orbit (some points along the orbit are marked 
with different symbols)
; the 
cross indicates the Galaxy center. 2: distance in kpc of the GC from the 
Galaxy center, as a function of time. 
The 
different symbols 
correspond to that in 
the previous panel. 3: GC orbital angular velocity in {\it rad/Myr},
as a function of time. 4: angular acceleration in {\it rad/$Myr^2$}.5:
solid curve: angle $\alpha$ formed by the inner part of the tails and the galactic center 
direction, vs time; 
dashed curve: angle $\beta$ formed by the inner part
of the tails and the cluster velocity, vs time. Both angles are in degrees.
\label{tail-radI}}
\end{figure}

\begin{figure}
\includegraphics[angle=270,scale=0.7]{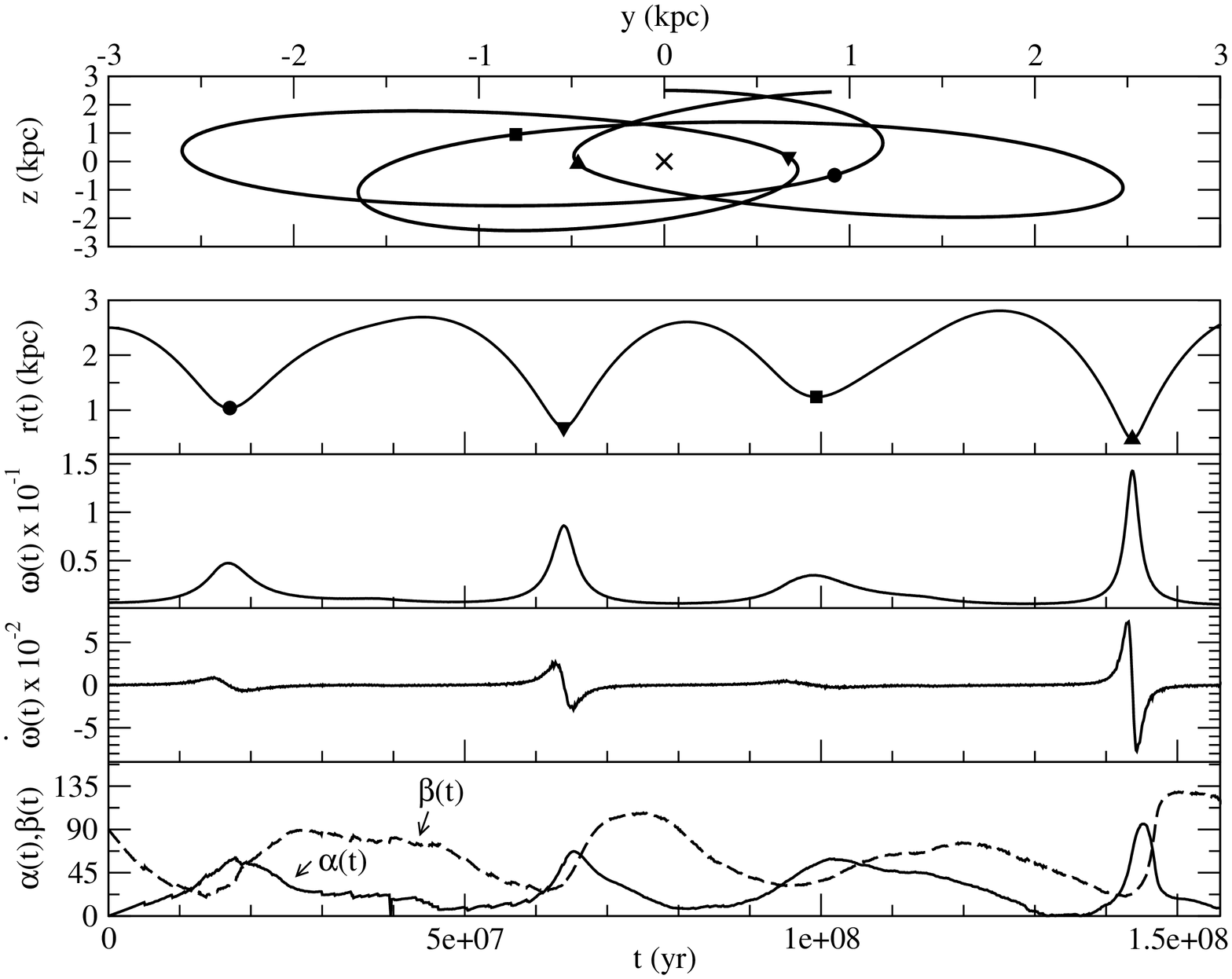}
\caption{Same as Fig.\ref{tail-radI}, but for the GC moving on orbit II.
\label{tail-radII}}
\end{figure}

\begin{figure}
\includegraphics[angle=270,scale=0.7]{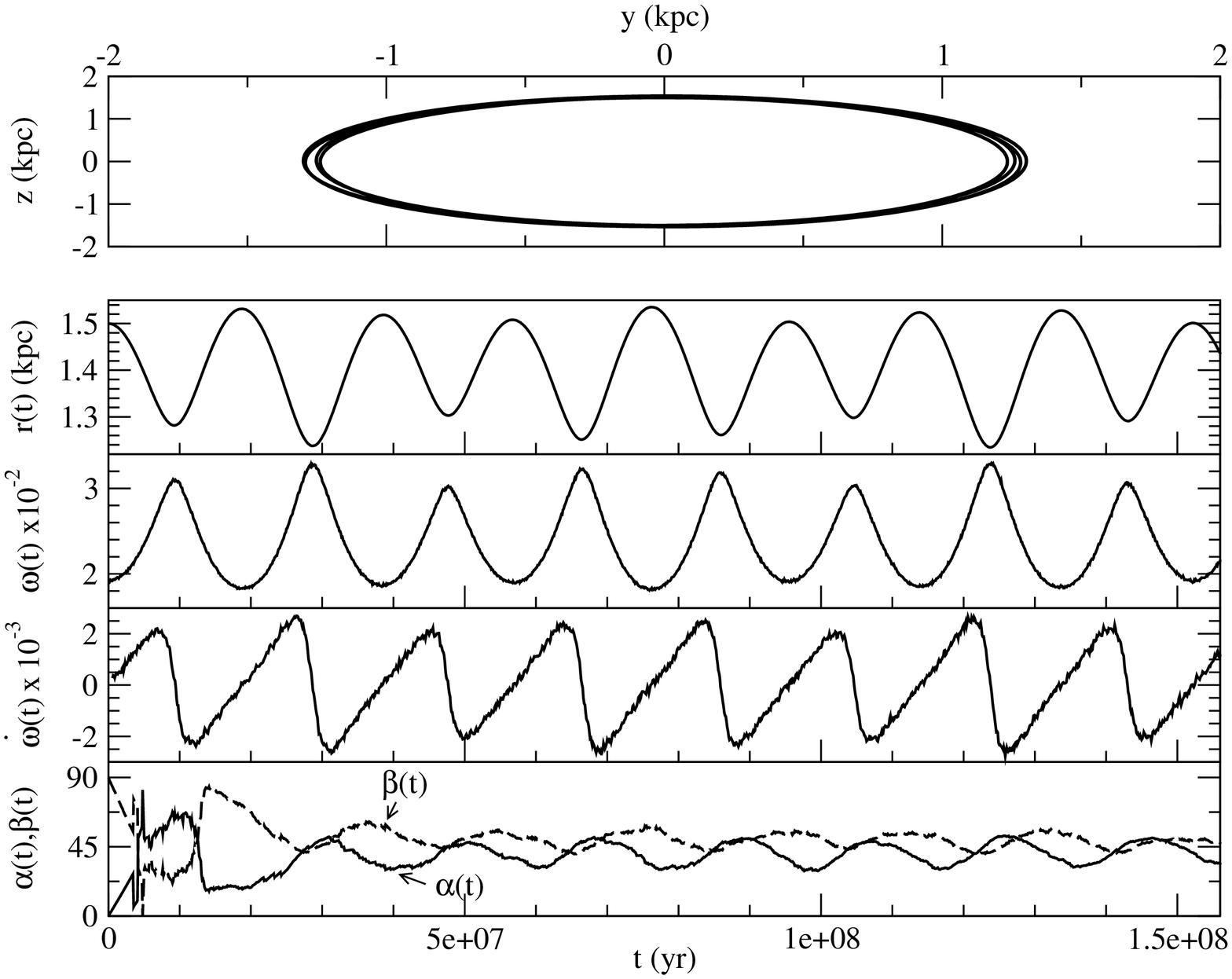}
\caption{Same as Fig.\ref{tail-radI}, but for the GC moving on orbit III.
\label{tail-radIII}}
\end{figure}

In Figure \ref{tail-radI}, which refers to the orbit I, $\alpha$ is represented by two,
barely distinguishable, solid curves, corresponding respectively to the GC 
with $c=1.02$ and $c=0.8$.
The first result is that the time evolution of $\alpha$ is independent 
of the GC concentration (in the interval of $c$ studied);
this means that the morphology and orientation of the tails depend on GC orbit,
rather than on GC internal parameters. \\ Moreover, the tail is elongated toward 
the galactic center
for most of the time: more in particular, if in Fig.\ref{tail-radI}
we exclude the regions around the pericenter, the average value of
$\alpha$ is $\sim 10^\circ$, i.e. the inner tail has 
only a slight deviation respect to the galactic center.

When approaching the pericenter, instead, the Coriolis and
the $\dot{\mbox{\boldmath{$\omega$}}}$ terms (in
Eq.\ref{equat}) are parallel and grow up. This induces
a rapid alignement of the tails with the GC orbit.
Beyond the pericenter, the Coriolis acceleration decreases and
$\dot{\mbox{\boldmath{$\omega$}}}$
becomes antiparallel to the former. The net effect yields a
realignment of the tails with the direction of the galactic
center. Figure \ref{tail-radII}, which refers to orbit II, 
confirms the results found for the previous orbit.

Finally, Figure \ref{tail-radIII} shows the case of the quasi-circular orbit (orbit III).
Along this orbit, both the distance of the GC from the galactic center and
the orbital angular velocity are nearly constant. This leads to a roughly constant $\alpha\sim
40 ^\circ$.
This suggests that the change in time of the orbital angular velocity
is what determine the tails orientation.

\subsubsection{Tail-orbit alignement}\label{tailvel}

The dashed curve shown in the bottom panel of Figures 
\ref{tail-radI},\ref{tail-radII} and \ref{tail-radIII} represents the
angle $\beta$, previously defined.
In the case of eccentric orbits,
the first striking feature is that when approaching the pericenter the angle $\beta$ 
decreases, reaches a minimum at the pericenter and then increases moving away from the pericenter.
The maximum value of $\beta$ is reached between the pericenter and the apocenter, depending on 
the shape of the GC orbit. However, in correspondence of the apocenter, 
$\beta$ is $\approx 90^\circ$ in agreement with the value $\alpha \approx 0^\circ$.
At the apocenter the tail is aligned along the radial direction and roughly perpendicular to the 
GC velocity.

It is worth noting that, in all the cases considered here, the tidal tails deviate
considerably by the GC velocity direction; only for very eccentric
orbits and close to the pericenter, the angle $\beta$ reaches a minimum value of
$\approx 17^\circ$. 
This indicates that the extrapolation of the cluster orbit from the
elongation of this part of the tidal streams leads, in general, to
predict GC paths with large errors.

 


\section{Conclusions}\label{concl}

This work is devoted to the study of the morphology and orientation of tidal 
tails surrounding globular clusters, in order to understand to what extent they 
trace the GC orbit and if some correlations exist between their orientation and 
the GC orbital phase.\\
The main findings of our work can be resumed as follows:
\begin{enumerate}
\item Tidal tails are good tracers of GC orbit only on the large scale, being the 
{\it inner part of the streams never aligned with the GC path}, unless the system
moves on very eccentric orbit and close to the orbital pericenter.
\item A {\it strong correlation} exists between the orientation of the inner tails and 
GC orbital position: they tend to be more elongated along the GC orbit when the 
cluster approaches the pericenter, while, in turn, they tend to align toward the
galactic center when the GC approaches the apocenter.
\item We have shown that a key role in determining the tails morphology is
played by the orbital angular velocity of the system and its variation
with time. 
In the case of the least eccentric orbit, a nearly constant angular velocity 
determines an almost constant orientation of tails with respect to the cluster orbit 
and the galactic center direction. 
The amplitude of $\dot{\mbox{\boldmath{$\omega$}}}$ 
contributes, too, to the alignement
of tails with the GC path, in particular just beyond the pericenter. 
\item The existing correlation between tails elongation and GC
position along its orbit
can be easily understood when referring to a non-inertial frame centered on the GC center,
due to the role played by non inertial accelerations, as the Coriolis
acceleration and that related to $\dot{\mbox{\boldmath{$\omega$}}}\times \bf {r}'_i$.
\end{enumerate}

We want to outline that our findings are in good agreement with some
observational results.
In particular we refer to the galactic globular clusters sample
studied by \citet{lmc00}, considering the 7 clusters
for which the
reliability level of the observed tidal streams is highest and
orbital parameters are available \citep{dinescu}.
Six clusters of this subsample \citep[see][Table 5]{lmc00}
have tidal tails clearly elongated along the galactic center direction,
and eccentricities greater than 0.62. This in
agreement with our results that GCs, on eccentric orbits, have tidal tails elongated
towards the galactic center for most of their path.
The only exception is represented by NGC5139 ($\omega$ Cen),
which has tidal tails deviating from the galactocentric
direction and extending towards the
galactic plane; this situation does not contradict our results
considering that NGC5139 is presently
only 1.3 kpc above the galactic disk and is most probably 
undergoing a strong disk-shocking.\\

The simulations also reproduce the formation of multiple tails
around globular clusters, as observed for NGC288 \citep{lmc00} and for the galactic 
satellite Willman 1 \citep{wil}. 
Our simulations show that these features are expected for GCs on eccentric orbits 
near their apogalacticon, while they are absent in the case of GCs moving on 
less eccentric orbits.
Actually, their formation can be explained by the behaviour of the Coriolis acceleration 
along the orbit: when its value is large, stars escaping from the cluster are pushed toward 
the orbital path.
This is particularly evident around the pericenter; at the apocenter, where 
the Coriolis acceleration is weaker, stars tend to escape from the cluster 
more radially. 
At every pericenter passage, this effect produce a new tidal tail roughly
aligned to the cluster path.

\section{Acknowledgements}
The authors acknowledge the Centro de Astrobiologi\'a (Madrid, Spain) for the use of 
its computational facilities. We also thank prof. D. Heggie for useful comments and 
for pointing us an error in one of the figure catpions.

\appendix
\section{Equations of motion in the non-inertial reference frame}\label{app}
Using a no-inertial reference frame $(x^{\prime},y^{\prime},z^{\prime})$ with the origin in the cluster center-of-mass, with the $(x^{\prime},y^{\prime})$ plane coinciding with the cluster orbital plane and $x^{\prime}$ always pointing to the galactic center (so that this reference frame rotates with the GC angular velocity  $\mbox{\boldmath{$\omega$}}$ respect to the inertial reference system  $(x,y,z)$), the position of the $i-th$ star in the galactocentric $(x,y,z)$ reference frame can be expressed simply as:
\begin{equation}
{\bf r}_i={\bf r}_{GC}+{\bf r'}_i
\end{equation}
being ${\bf r}_{GC}$ the position of the cluster center-of-mass in the inertial system.
Rewriting the previous equation in terms of the $(x^{\prime},y^{\prime})$ components (we omit the $z'$ one for simplicity) we have:
\begin{equation}\label{pos}
x_i \mbox{\boldmath{$\hat{x}$}}+y_i \mbox{\boldmath{$\hat{y}$}}=x_{GC}\mbox{\boldmath{$\hat{x}$}}+y_{GC}\mbox{\boldmath{$\hat{y}$}}+x'_i \mbox{\boldmath{$\hat{x'}$}}+y'_i \mbox{\boldmath{$\hat{y'}$}}
\end{equation}
being $\mbox{\boldmath{$\hat{x}$}}$,$\mbox{\boldmath{$\hat{y}$}}$ and $\mbox{\boldmath{$\hat{x'}$}}$,$\mbox{\boldmath{$\hat{y'}$}}$ the unity vectors in the two frames, respectively.
Derivating Eq.\ref{pos} with respect to time $t$, one obtains:
\begin{eqnarray}\label{vel}
\dot{x}_i \mbox{\boldmath{$\hat{x}$}}+\dot{y}_i \mbox{\boldmath{$\hat{y}$}}&=&\dot{x}_{GC}\mbox{\boldmath{$\hat{x}$}}+\dot{y}_{GC}\mbox{\boldmath{$\hat{y}$}}+\dot{x'}_i\mbox{\boldmath{$\hat{x'}$}}+\dot{y'}_i\mbox{\boldmath{$\hat{y'}$}}+x'_i\mbox{\boldmath{$\dot{\hat{x}'}$}}+y'_i\mbox{\boldmath{$\dot{\hat{y}'}$}}=\\
& &\dot{x}_{GC}\mbox{\boldmath{$\hat{x}$}}+\dot{y}_{GC}\mbox{\boldmath{$\hat{y}$}}+\dot{x'}_i\mbox{\boldmath{$\hat{x'}$}}+\dot{y'}_i\mbox{\boldmath{$\hat{y'}$}}+\mbox{\boldmath{$\omega$}}\times(x'_i \mbox{\boldmath{$\hat{x'}$}}+y'_i \mbox{\boldmath{$\hat{y'}$}})
\end{eqnarray}
being
\begin{eqnarray}
\mbox{\boldmath{$\dot{\hat{x}'}$}}&=&\mbox{\boldmath{$\omega$}}\times\mbox{\boldmath{$\hat{x'}$}}\label{xdot}\\
\mbox{\boldmath{$\dot{\hat{y}'}$}}&=&\mbox{\boldmath{$\omega$}}\times\mbox{\boldmath{$\hat{y'}$}}\label{ydot}
\end{eqnarray}
Derivating Eq.\ref{vel} with respect to time $t$ and using Eqs. \ref{xdot}-\ref{ydot}, one obtains:
\begin{eqnarray}
\ddot{x}_i \mbox{\boldmath{$\hat{x}$}}+\ddot{y}_i \mbox{\boldmath{$\hat{y}$}}&=&\ddot{x}_{GC}\mbox{\boldmath{$\hat{x}$}}+\ddot{y}_{GC}\mbox{\boldmath{$\hat{y}$}}+\ddot{x'}_i \mbox{\boldmath{$\hat{x'}$}}+\ddot{y'}_i \mbox{\boldmath{$\hat{y'}$}}+2\mbox{\boldmath{$\omega$}}\times(\dot{x'}_i\mbox{\boldmath{$\hat{x'}$}}+\dot{y'}_i\mbox{\boldmath{$\hat{y'}$}})+\\
& &\mbox{\boldmath{$\omega$}}\times\left(\mbox{\boldmath{$\omega$}}\times(x'_i \mbox{\boldmath{$\hat{x'}$}}+y'_i \mbox{\boldmath{$\hat{y'}$}})\right)+\dot{\mbox{\boldmath{$\omega$}}}\times(x'_i \mbox{\boldmath{$\hat{x'}$}}+y'_i \mbox{\boldmath{$\hat{y'}$}})
\end{eqnarray}
and so, finally, Eq.\ref{equat}.



\clearpage

\clearpage

\clearpage

\begin{deluxetable}{rcc}
\tablecaption{Parameters for the Galactic model \citep{allsant91}\label{tbl1}}
\tablewidth{0pt}
\startdata
Bulge & $M_B$ & $1.41 \times 10^{10} M_{\odot}$\\
 & $b_B$ & 387.3 pc\\
Disk & $M_D$& $8.56\times 10^{10} M_{\odot}$\\
& $a_D$ & 5317.8 pc\\
& $b_D$ & 250.00 pc\\
Halo & $M_H$ &$10.7\times10^{10} M_{\odot}$\\
& $a_H$& 12000 pc
\enddata
\end{deluxetable}

\begin{deluxetable}{cccccc}
\tablecaption{GC structural parameters \label{tbl2}}
\tablewidth{0pt}
\startdata
$M_{GC}[M_{\odot}]$ & $r_c$ [pc] & $r_t$ [pc] & $c=log(r_t/r_c)$ & $\sigma$[km/s] 
\\
& & & & &\\
$4.7 \times 10^5$ &2.2 & 23.5 & 1.03 & 10.5\\
$4.7 \times 10^5$ &3.5 & 22.1 & 0.81 & 10.5
\enddata
\tablecomments{Col. (1): cluster total mass. Col. (2-4): core radius, cutoff radius
and concentration parameter of the King's model. Col. (5): central velocity dispersion.}

\end{deluxetable}

\begin{deluxetable}{cccccccc}
\tablecaption{GC initial orbital parameters \label{tbl3}. }
\tablewidth{0pt}
\startdata
Orbit ID & $x$[pc] & $y$[pc] & $z$[pc] & $v_x$[km/s] & $v_y$[km/s] & $v_z$[km/s] & $e$ 
\\
& & & & & & &\\
I &0. & 0. & 4000. & 0. &  53.3 & 0. & 0.8\\
II & 0. & 0. & 2500. & 0. & 116.7 & 0. & 0.7\\
III & 0. & 0. & 1500. & 0. & 200.0 & 0. & 0.1
\enddata
\tablecomments{Col. (1): orbit identification. Col. (2-7): cluster initial center-of-mass position
and velocity components. Col (8): orbit eccentricity [Eq.(\ref{ecc})] }
\end{deluxetable}


\end{document}